\documentclass[twocolumn,showpacs,prb]{revtex4}

\usepackage[dvips]{graphicx}

\begin{document}

\title{Correlated Photon-Pair Emission from a Charged Single Quantum Dot}

\author{S.~M.~Ulrich$^{1}$\cite{ADDRESS}, M.~Benyoucef$^{1}$, P.~Michler$^{1}$, N.~Baer$^{2}$, P.~Gartner$^{2,3}$, F.~Jahnke$^{2}$, M.~Schwab$^{4}$, H.~Kurtze$^{4}$, M.~Bayer$^{4}$, S.~Fafard$^{5}$, and Z.~Wasilewski$^{5}$}

\affiliation{$^{1}$5$^{th}$ Physics Institute, University of Stuttgart, 70550 Stuttgart, Germany}
\affiliation{$^{2}$Institute of Theoretical Physics, University of Bremen, P.O. Box 330440, 28334 Bremen, Germany}
\affiliation{$^{3}$National Institute of Materials Physics, P.O. Box MG-7, Bucharest-Magurele, Romania}
\affiliation{$^{4}$Experimental Physics II, University of Dortmund, 44227 Dortmund, Germany,}
\affiliation{$^{5}$Institute for Microstructural Sciences, National Research Council of Canada, Ontario K1A 0R6, Canada}

\begin{abstract}

The optical creation and recombination of charged biexciton and trion complexes in an (In,Ga)As/GaAs quantum dot is investigated by micro-photoluminescence spectroscopy. Photon cross-correlation measurements demonstrate the temporally correlated decay of charged biexciton and trion states. Our calculations provide strong evidence for radiative decay from the excited trion state which allows for a deeper insight into the spin configurations and their dynamics in these systems.            

\end{abstract}


\maketitle

In the expanding research fields of quantum information technology the achievement of deterministic, i.e. fully controllable single-photon devices is of highest interest as this provides the key for the realization of novel scalable data processing schemes in quantum computing and cryptography\cite{D.Bouwmeester:2000}. In recent years, various types of semiconductor-based quantum dot structures have been demonstrated as promising concepts\cite{P.Michler:2000c, C.Santori:2001b, E.Moreau:2001b, C.Santori:2002b, K.Sebald:2002, V.Zwiller:2003} where especially the development of electrically driven sub-Poisson photon sources\cite{O.Benson:2000, Z.Yuan:2002} marks an important step.  

As was first shown by Moreau et al.\cite{E.Moreau:2001a}, single QDs are also capable to serve as sources of temporally correlated photon pairs formed by the cascaded decay of biexcitonic and excitonic states. More detailed investigations have revealed a strong linear polarization correlation of those photon pairs \cite{C.Santori:2002, R.M.Stevenson:2002, S.M.Ulrich:2003} in consistence with a QD asymmetry-induced lift of the exciton states' degeneracy\cite{M.Bayer:1999}.

Besides studies on the creation and decay of neutral (multi-)excitonic carrier configurations in QDs, in recent investigations also the \textit{charged} counterparts of these complexes have attracted increasing interest \cite{A.Wojs:1997, F.Findeis:2001, J.J.Finley:2001, L.Besombes:2003, B.Urbaszek:2003} due to their possible applications in quantum information processing schemes. In particular, magnetoluminescence measurements have enabled to study the QD trionic fine structure\cite{M.Bayer:2002, I.A.Akimov:2002} in high detail. Further investigations have also demonstrated coherently manipulated charged QD states\cite{L.Besombes:2003}.  

In this paper we provide direct evidence for the cascaded emission from a charged biexciton via an excited trion state which is supported by a detailed theoretical analysis. In addition, our studies enable a deeper insight into the fundamental electronic properties of charged QDs and the corresponding optical transitions.    

The QD sample under investigation was grown by molecular beam epitaxy (MBE) using $n$-doped (001)-oriented GaAs as the substrate material. On top of a thin GaAs buffer layer a single layer of self-assembled InAs islands was deposited which formed 3-D confined individual quantum dots (height: 1-2~nm; width: 15-20~nm) within a thin ternary wetting layer (1~ML) after final capping by a 50~nm top layer of GaAs. The QD surface density was $\approx 10^{10}$~cm$^{-2}$. In order to enable studies on individual QDs, an array of single mesa structures was fabricated in a post-growth step by combined electron beam lithography and wet chemical etching. The investigations discussed in the following were performed on a selected 320~nm diameter mesa.

Our experiments have been performed on a combined low-temperature (4~K) micro-photoluminescence ($\mu$-PL) system and a Hanbury Brown and Twiss (HBT)-type photon correlation setup for investigations on the photon emission statistics (for details see Ref.[\onlinecite{S.M.Ulrich:2003}]). Spectral selection within our HBT setup was achieved by the use of tunable acousto-optical ($\Delta\lambda=1$~nm) and/or narrow-band interference filters ($\Delta\lambda=0.5$~nm) inside the detection paths. The histograms $n(\tau)$ (normalized: $g^{(2)}(\tau)$) of photon correlation events with delay times $\tau$ were measured by a multi-channel-analyser. For quasi-resonant (p-shell) or off-resonant (above GaAs barrier) continuous wave (cw) excitation a Ti:Sapphire laser was used. For pulsed auto-correlation measurements, the laser system was used in its picosecond mode ($\Delta t_{pulse} \approx 2$~ps at $76.2$~MHz). A steep angle ($30^{\circ}$) optical fiber geometry was used for excitation which allowed for focusing down to a spot diameter of $\sim$~10~$\mu$m on the sample surface. 

\begin{figure}
\begin{center}
\includegraphics[width=9.1cm]{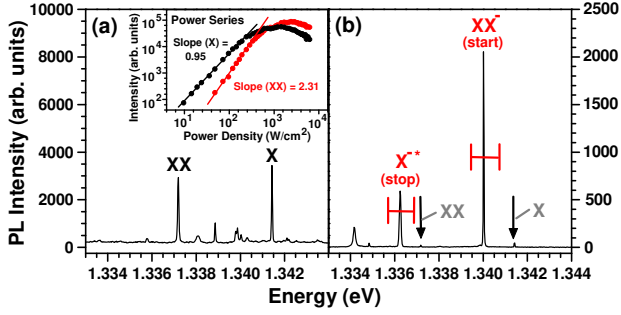}
\caption{(Color online) (a) $\mu$-PL spectra from a single (In,Ga)As QD as observed for off-resonant (i.e. above barrier) cw laser excitation. Inset: Power dependence reflecting the onset and saturation behaviour of excitonic (X) and biexcitonic (XX) emission. (b) Spectra taken on the same QD under quasi-resonant (p-shell) excitation revealing PL lines being addressed to charged complexes. Horizontal markers correspond to the filter bandpass in photon correlation experiments.\label{FIG.1}}
\end{center}
\end{figure}

Fig.~\ref{FIG.1}(a) shows $\mu$-PL spectra as obtained for above-barrier cw excitation 
($E_{exc}$=~1.597~eV; $P_{exc}$=~1.9~kW/cm$^{2}$). Apart from an almost unstructured small PL background the spectral window of interest is dominated by a pair of intense narrow lines at 1.3414~eV and 1.3372~eV which exhibit nearly resolution-limited linewidths of $55 \pm 10\,\mu$eV and $75 \pm 10\,\mu$eV, respectively. As is depicted in the inset of Fig.~\ref{FIG.1}(a), the PL of the two lines reveals a linear (slope~=~$0.95 \pm 0.05$) and super-linear (slope~=~$2.31 \pm 0.05$) increase of intensity with excitation density over almost two decades. Together with the observed relation between onset and saturation this reflects the excitonic (X) and biexcitonic (XX) origin of these emission lines. Additional evidence is obtained from time-resolved spectroscopy providing a radiative lifetime $\tau_{X} \approx 1.0$~ns and $\tau_{XX} \approx 500$~ps for these lines. The assignment of both peaks to the same QD is strongly supported by the observed characteristic line spacing of $\Delta E_{X-XX} = 4.2\pm0.1$~meV which is consistent with the typical biexciton-exciton Coulomb exchange energy for those dots.

As is shown in Fig.~\ref{FIG.1}(b), the emission spectrum completely changes when the exciting laser is tuned energetically into the QD p-shell ($E_{exc}^{res} = 1.4068\pm0.0004$~eV; $P_{exc}= 1.9\,\pm 0.1$~kW/cm$^{2}$). This quasi-resonant excitation reveals the onset of a new set of almost background-free narrow lines centered at 1.3400~eV ($50 \pm  5\,\mu$eV), 1.3362~eV ($80 \pm 5\,\mu$eV) and 1.3342 eV ($110 \pm 5\,\mu$eV) on the low energy side of the strongly suppressed X and XX PL signals (marked by arrows). In the following, we focus on the two former lines which correspond to the radiative decay of the \textit{charged biexciton} complex ($1e^{2}2e^{1}1h^{2} \mapsto 1e^{1}2e^{1}1h^{1}$) and the consecutive \textit{trion} recombination from its \textit{excited} state ($1e^{1}2e^{1}1h^{1} \mapsto 2e^{1}$), respectively, as will be demonstrated below. 

\begin{figure}
\begin{center}
\includegraphics[width=9.1cm]{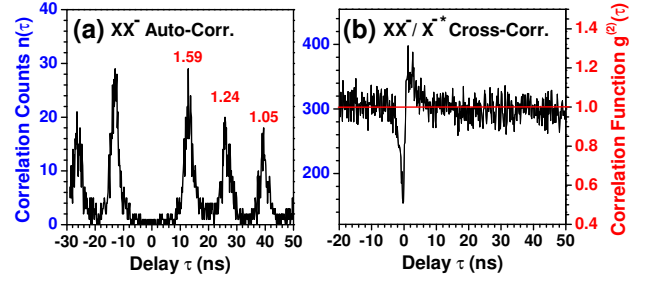}
\caption{(Color online) (a) Photon auto-correlation measurement under ps pulsed quasi-resonant excitation ($E_{exc} = 1.407$~eV) as obtained from the 1.340~eV PL line (see Fig.~\ref{FIG.1}(b)). Numbers: Poisson-normalized peak areas. (b) Cross-correlation trace obtained from the 1.3400~eV (XX$^{-}$: "start") and 1.3362~eV (X$^{-\ast}$: "stop") PL lines in Fig.~\ref{FIG.1}(b). \label{FIG.2}}
\end{center}
\end{figure}

In order to investigate the photon emission statistics of the observed PL lines we have performed photon auto-correlation measurements using the HBT setup. As an example, the result obtained from the intense XX$^{-}$ peak (1.3400 eV) under quasi-resonant pulsed excitation is shown in Fig.~\ref{FIG.2}(a). Multiple-photon ($\geq 2$) emission is found to be fully suppressed on this PL line which manifests itself in the absence of the zero delay ($\tau = 0$~ns) coincidence peak. This behaviour demonstrates almost perfect (background-free) triggered single-photon generation from an individual QD.

Furthermore, in Fig.~\ref{FIG.2}(a) the correlation peaks at delay times $\tau = n\,T_{laser}$ ($n = \pm 1, \pm 2,...$) corresponding to correlation events of photons following subsequent laser excitation cycles $T_{laser} = 13.12$~ns reveal a clear \textit{bunching}-like behaviour. This means, we observe a higher probability of detecting another XX$^{-}$ photon in the next (previous) excitation cycle than in later (earlier) cycles. Such an effect has been reported recently on similar InAs QDs especially under the condition of \textit{pure quasi-resonant} pumping into an excited state \cite{C.Santori:2002b, C.Santori:2004} and is interpreted as a fast "blinking" process between the neutral and charged state configuration of a single QD. Therefore, the blinking is assumed to reflect the condition of a favored even or odd-number carrier occupancy of the QD. Furthermore, it is important to note that this strongly suggests a residual charging of the QD by ionization of adjacent impurities \cite{M.Sugisaki:2002} which supports the preliminary assignment of the PL signatures given in Fig.~\ref{FIG.1}(b). Applying a fit formula \cite{C.Santori:2004} of the type $g^{2}(\tau) = 1 + g_{1} \cdot exp(-\tau/\tau_{blink})$ for discrete values $\tau = n\,T_{laser}$ ($n = \pm 1, \pm 2,...$) to the Poisson-normalized correlation peak areas of the data in Fig.~\ref{FIG.2}(a), we can extract the parameters $g_{1}=1.67 \pm 0.14$ for the bunching amplitude and $\tau_{blink} = 12.8 \pm 2.0$~ns for the blinking time constant. The short timescale of this effect is consistent with the generally expected behaviour for a regime of excitation close to the saturation level \cite{C.Santori:2004}. 

The assignment of the observed PL under quasi-resonant excitation to charged complexes is supported by a simple model scheme: In the absence of laser excitation a (doping-related or equivalently intrinsic, e.g. donor-type) impurity in the vicinity of an individual QD enables an excess electron to relax into the QD ground state thus initially leaving a charged dot and a nearby ionized impurity. Consequently, under p-shell excitation the creation of odd-number charge states (i.e., e-h pairs plus an extra electron) should be favored. In contrast to this, for above-barrier pumping the effect of \textit{photo-depletion} \cite{A.S.Chaves:1995, A.Hartmann:2000} is expected: The dissociation of hot e-h pairs through the local QD-impurity Coulomb field leads to the attraction of holes into the QD whereas excess electrons are also trapped by the impurity centers. In this case, the radiative decay of neutral X and XX complexes is predicted.

In addition, we have performed photon \textit{cross}-correlation measurements under quasi-resonant continuous wave (cw) excitation in order to identify the temporal order for the emission of the discussed XX$^{-}$ and X$^{-\ast}$ lines. Fig.~\ref{FIG.2}(b) shows the corresponding correlation histogram as measured $n(\tau)$ (left axis) and in the Poisson-normalized $g^{(2)}(\tau)$ (right axis) representation. Using the XX$^{-}$ emission line (1.3400~eV) as the "start" and the X$^{-\ast}$ line (1.3362~eV) as the "stop" trigger of the HBT, a pronounced signal asymmetry was observed in the vicinity of zero delay. The anti-bunching for $\tau < 0$ together with the observed bunching behaviour at positive delays $\tau > 0$ clearly identifies the two decay channels under investigation to be temporally correlated, i.e. to form a \textit{radiative cascade} originating from the same QD.

For a theoretical analysis of the experimental data discussed above the eigenstates and the emission spectra of the QD have been computed using a full configuration interaction (FCI) procedure~\cite{FCI,BGJ}. These results indeed show that the X$^{-*}$ emission line appears at a significantly lower energy than the XX$^-$ line, while the emission from the trionic \textit{ground state} (i.e., X$^{-}$) is placed at slightly higher energies and therefore is not a good candidate.    

\begin{figure}[!htb]
\begin{center}
\includegraphics[width=5.4cm]{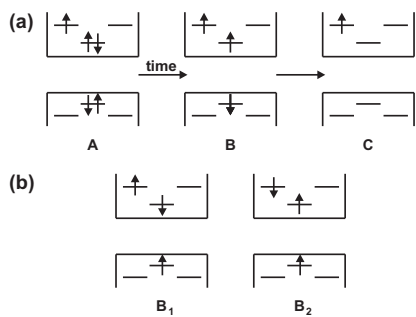}
\caption{(a) Scheme of the $XX^- \rightarrow X^{-*} \rightarrow e$ cascade: Starting from the charged biexciton ground state $A$ an excited charged exciton state $B$ is reached by a e-h-recombination process. The second recombination process leads to an excited dot state $C$ with only one electron. (b) Trion configurations with opposite spins for the $s$ and $p$ electrons. \label{FIG.3}}
\end{center}   
\end{figure}

The charged biexciton PL line is brought below the trionic ground state emission line by the (charged) biexciton binding energy. In the case of the \textit{excited} trionic line the exchange interaction between the $s-$shell and $p-$shell carriers overcomes the biexcitonic binding energy and reverses the order of the lines. This can be understood by considering the diagrams in Fig.~\ref{FIG.3}(a). One $s$-level and two $p$-levels are present in each band. All levels are spin degenerate and, assuming cylindrical symmetry for the QD in the wetting-layer plane, the two $p$-levels are degenerate, too. 

The eigenstates provided by the FCI calculations are combinations of many configurations. In Fig.~\ref{FIG.3}(a) only the dominant configurations of the states taking part in the cascade are shown. Their high participation ratio ($>96\%$) allows us to estimate the corresponding energies according to the approximation scheme discussed in Ref.[\onlinecite{BGJ}]. For the energy of the initial configuration $A$ this leads to $E_A = 2E^{X}_s + \varepsilon^e_p-X_{sp}$, i.e. the sum of two $s-$exciton energies $E^{X}_s$, the one-particle $p-$shell electronic energy $\varepsilon^e_p$ and the (attractive) exchange energy $X_{sp}$ between $s$ and $p$ electrons with parallel spin. Similarly, $E_B = E^{X}_s + \varepsilon^e_p-X_{sp}$ and $E_C = \varepsilon^e_p$. For the recombination of the first electron-hole pair, in fact two possible decay paths exist which lead to a final charged exciton state including either (1) two electrons with \textit{identical} spin orientations (configuration $B$ in Fig.~\ref{FIG.3}(a)) or (2) two electrons with \textit{opposite} spins. On this latter configuration (Fig.\ref{FIG.3}(b)) we will comment below, here we first discuss the first decay channel.

Strictly speaking, the additivity of different energy contributions leaves out some configuration-interaction effects. One of them is the biexcitonic binding energy, which can be accounted for only by the interaction of the configurations of Fig.~\ref{FIG.3}(a) with the other possible arrangements of
the carriers on the available states. Considering only the $s$- and $p$-shells and with the material parameters of Ref.[\onlinecite{BGJ}] one obtains a biexciton binding energy of $\approx$~2~meV for the neutral biexciton and $\approx$~1.2~meV for the charged one. Therefore, with the main configuration-interaction effect included, $E_A$ should read $E_A = 2E^{X}_s - \Delta + \varepsilon^e_p-X_{sp}$, with $\Delta$ the (charged) biexciton binding energy. As a result one finds the following values for the emission lines: 
\begin{eqnarray}
E_{XX^{-}} &=& E_A-E_B = E^{X}_s  - \Delta \nonumber \\
E_{X^{-*}}&=& E_B-E_C = E^{X}_s - X_{sp}\label{eq:engy}. 
\end{eqnarray} 
The difference in the positions of the two lines is given by $X_{sp}-\Delta$. The exchange integral $X_{sp}$ (in our case $\approx4.9$~meV) is significantly larger than the biexcitonic binding energy $\Delta$, which explains the energetic positions $E_{X^{-*}}<E_{XX^{-}}$.   

To complete the discussion, we point out that the initial charged biexciton state is actually fourfold degenerate, since four states are available to the $p$-electron. Moreover, the $s$-electron left in the excited trion may not necessarily have the same spin as the $p$-electron. Such a situation is shown in the configuration $B_1$ of Fig.~\ref{FIG.3}(b). In this case the above argument seems to fail, since the exchange integral apparently plays no role any more. Though, the FCI calculation shows that this is \textit{not} the case. Actually, the configuration $B_1$ alone is not an eigenstate of the problem. It gets mixed with the configuration $B_2$ (which cannot be directly obtained from the radiative decay of $A$) and the resulting state has exactly the same energy as the state $B$. This is by no means a coincidence: The electrons in the configuration $B$ have the spin state $S^e=1, S_z^e=1$ and the sum of $B_1$ and $B_2$ corresponds to $S^e=1, S_z^e=0$. These states are necessarily degenerate in an eigenvalue problem which conserves both the electron and the hole spin. 

In addition to the experiments discussed before, we have also tested the possibility to generate polarization-correlated photon pairs by the XX$^{-}\mapsto$~X$^{-*}$ cascade under investigation. Surprisingly, within our experimental accuracy no polarization correlations between the two emitted photons could be observed. Together with the blocked relaxation of the $p$-shell electron in the excited trion configuration (see preceding discussion), these data allow us to get some insight into the spin dynamics in these systems. For higher dimensional systems it is known (see, e.g. Ref.~[\onlinecite{X.Marie:2000}]) that the fastest relaxation occurs for the \textit{hole} spin, followed by that of the \textit{exciton}. It takes the longest time for the \textit{electron} spin to flip. This hierarchy seems to be maintained also for the QD structures under study: The fact that emission from the excited trion complex is observed indicates a considerably longer spin-flip time for the $p$-shell electron  than for the electron-hole recombination. Since time-resolved PL on single QDs revealed a decay time of $\approx$~1~ns, this gives a lower limit for the electron spin-flip time in the excited state. On the other hand, the missing photon polarization correlation~\cite{A.Kiraz:2002} in the cascade can be explained only under the assumption of an $s$-shell exciton spin flip being much faster than the radiative lifetime (1~ns). 

In conclusion, the emission from charged exciton and biexciton carrier complexes have been observed under pure quasi-resonant excitation. Besides the capability of background-free triggered single-photon generation, the cascaded nature of these decay channels was directly demonstrated by means of photon cross-correlation measurements. From a direct comparison with many-particle QD eigenstate calculations, especially for the involved trion complex clear indications for a predominant recombination from its excited state was revealed, thus reflecting a long-lived excess electron spin configuration.   

The work was supported by the Volkswagenstiftung (I/76 142) and the Deutsche Forschungsgemeinschaft through the {\it Quantum Optics in Semiconductor Nanostructures} research group (FOR 485/1-1). The Bremen group acknowledges a grant for CPU time at the NIC J\"{u}lich. We also acknowledge the support by the NRC-Helmholtz S\&T cooperation initiative. Furthermore, the authors would like to thank P.~Hawrylak for helpful discussions.

\end{document}